\documentstyle[12pt]{article}
\def\erf{\mathop{\rm erf}\nolimits}

\def\Re{\mathop{\rm Re}\nolimits}

\begin{document}


\centerline{\huge Examples of potentials with  convergent}
\centerline{\huge Schwinger --- DeWitt expansion}
\vspace{1cm}
\centerline{{\Large V.~A.~Slobodenyuk}
                             \footnote{Physical--Technical Department,
                             Ulyanovsk State University,
                             432700 Ulyanovsk, Russian Federation.\\
                             E-mail: slob@themp.univ.simbirsk.su} }

\vspace{3cm}
{\large {\bf Running head:} Examples of potentials with  convergent
       Schwinger --- DeWitt expansion}

\vspace{4cm}
{\bf Address of author:} \\
{\it Physical--Technical Department} \\
{\it Ulyanovsk State University } \\
{\it L.Tostogo str. 42, 432700 Ulyanovsk} \\
{\it Russian Federation} \\
{\it Tel.: 7 (8422) 32-06-12} \\
{\it E-mail: slob@themp.univ.simbirsk.su}

\newpage

\begin{abstract}
Convergence of the Schwinger --- DeWitt expansion for the
evolution operator kernel for special class of potentials
is studied. It is shown, that this expansion,
which is in general case asymptotic, converges
for the potentials considered (widely used, in particular, in
one-dimensional many-body problems), and besides, convergence
takes place only for definite discrete values of the coupling
constant. For other values of the charge divergent expansion
determines the kernels having
essential singularity at origin (beyond usual $\delta$-function).
If one consider only this class of potentials
then one can avoid many problems, connected with asymptotic
expansions, and one get the theory with discrete values of the
coupling constant that is in correspondence with discreteness
of the charge in the nature.
This approach can be transmitted into the quantum field theory.
\end{abstract}

\medskip

\subsection*{Key words}

Evolution operator kernel, Schr\"odinger equation, convergence,
quantization of charge.

\bigskip

\sloppy

\section{Introduction}

In the quantum theory expansions in different parameters such as
the coupling constant~[Bender 1969, 1971~\cite{BW2}, 1971~\cite{BW3},
Lipatov 1977], the WKB--expansion, the
short-time Schwinger --- DeWitt expansion~[Schwinger 1951,
DeWitt 1965, 1975] the
perturbation expansion in phase-space technique~[Barvinsky 1995],
$1/n$-expansion~[Popov 1992], etc. are, as a rule, asymptotic.
This circumstance imposes essential restrictions on possibilities
of their using, makes the theory incomplete and compels one to look
for the ways of overcoming these restrictions
either by summation of divergent series with special methods (see,
e.g.,~[Kazakov 1980]), or by constructing new convergent
expansions~[Holliday 1980, Ushveridze 1983, Sissakian 1992],
 or by creating different approximate methods taking into
consideration the so-called nonperturbative effects.

Interesting information about possible way of overcoming the
problem of divergencies in some cases may be obtained from
investigation of the time dependence of the evolution
operator kernel with a help of the Schwinger --- DeWitt
expansion. For example, such investigations for a class
of potentials which is the family of bounded and
continuous functions that are formed from the Fourier
transforms of complex bounded measures were made in~[Osborn 1983].
An important feature of this expansion is that after
factorization of the contribution of the free kernel ("free" case
corresponds to $V\equiv 0$),
having at $t=0$ the singularity in the form of $\delta$-function
in space variables,
one can concentrate attention on rest part (denote it as $F$) which,
according to the initial condition, should be equal to 1 when $t=0$.
To understand behaviour at $t=0$ it is necessary to make analytical
continuation into the complex plain $t$.
When this continuation is made for the kernel,
then its analytical properties are masked by the singularity
which provides $\delta$-like behaviour in space variables.
But if the factorization of the free part of the kernel
is made, then the rest function can be continued into the entire
complex plain
$t$ and one can accurately examine its properties in the
neighborhood of origin.

If the Schwinger --- DeWitt expansion is convergent, then the
point $t=0$ is regular and initial condition is fulfilled in
rigorous sense. But if this expansion is divergent (note,
that usually it is treated as asymptotic~[Osborn 1983, Slobodenyuk 1995,
1996~\cite{MPLA2}]) then the
point $t=0$ is essential singular point for the function $F$.
In this case the initial condition may be fulfilled only in
asymptotic sense. The function $F$ tends to 1 when $t \to 0$
along the real positive semiaxis as continuous function,
but it is not analytic at $t=0$ and it does not have any
meaning at this point.

Nevertheless, it is enough to fulfill the initial
condition even in asymptotic sense that
unambiguous solution of the evolution problem to exist. So,
divergence of the Schwinger --- DeWitt expansion does not put
any formal restrictions on choose of the potentials
in the quantum theory. But using of the potentials  with
divergent expansion (if exact
solution is not known) is usually connected with problems of
different divergences (see, e.g., Sect. 4).

There exist possibility to avoid many of these problems in
some cases.
If one consider the potentials for which the Schwinger ---
DeWitt expansion converges, then one may get convergent
representation for the kernel and other physical values.
Such nontrivial potentials really exist. This paper is
devoted to consideration of some examples of such potentials
and to proving of convergence of the expansion for them.
These are the potentials
being widely used in one-dimensional many-body
problems~[Olshanetsky 1983, Calogero 1975, Sutherland 1971, 1972].
For definite discrete
values of the coupling constant the expansions for them are
convergent in the entire complex plain $t$. For other
values of the charge the expansions are asymptotic.
Phenomenon of existence of such potentials is very interesting.
Moreover, convergence of
the expansion only for discrete values of the coupling constant
may be connected with discreteness of the charge in the nature.

\section{The method of research}

The evolution operator kernel of the Schr\"odinger equation in
one-dimensional case is the solution of the problem
  \begin{equation} \label{f1}
  i\frac{\partial}{\partial t} \langle q',t\mid q,0 \rangle =
  -\frac{1}{2} \frac{\partial^2}{\partial q'^2}
  \langle q',t\mid q,0 \rangle + V(q') \langle q',t\mid q,0 \rangle,
  \end{equation}
  \begin{equation} \label{f2}
  \langle q',t=0\mid q,0 \rangle = \delta (q'-q).
  \end{equation}
Here and everywhere below dimensionless values, which are derived from
dimension ones in an obvious way, are used for the sake of
convenience.
The variable $t$ is treated as a complex one. If one means the proper
Schr\"odinger equation then $t$ is real.
We imply that $V(q)$ does not apparently depend on time.

As it is well known, in the free case ($V \equiv 0$) the solution of
the problem~(\ref{f1}), (\ref{f2}) is
  \begin{equation} \label{f3}
  \langle q',t\mid q,0 \rangle = \frac{1}{\sqrt{2\pi t}}
  \exp \left\{i \frac{(q'-q)^2}{2t} \right\} \equiv \phi(t;q',q).
  \end{equation}
The function $\phi$ has essential singularity at $t=0$, but this
singularity is such, that it provides the initial condition~(\ref{f2})
 to be fulfilled.

When interaction is present the kernel can be represented as
  \begin{equation} \label{f4}
  \langle q',t\mid q,0 \rangle = \frac{1}{\sqrt{2\pi it}}
  \exp \left\{i \frac{(q'-q)^2}{2t} \right\} F(t;q',q),
  \end{equation}
and besides, one can write for $F$ the expansion (the short-time
Schwinger--DeWitt expansion)
  \begin{equation} \label{f5}
  F(t;q',q) = \sum_{n=0}^{\infty} (it)^n a_n(q',q),
  \end{equation}
which, as a rule, is asymptotic and is usually utilized only
in that quality. We shall use representation~(\ref{f4}),
(\ref{f5}) to test the analytical properties of the evolution
operator kernel in variable $t$
and, in particular, ascertain its behavior for $t \to 0$.

For this purpose let us derive from~(\ref{f1}) the equation for $F$
  \begin{equation} \label{f6}
  i\frac{\partial F}{\partial t} =
  -\frac{1}{2} \frac{\partial ^2 F}{\partial q'^2} +
  \frac{q'-q}{it} \frac{\partial F}{\partial q'} + V(q')F.
  \end{equation}
Because the factorized function $\phi$ yet fulfils the initial
condition (\ref{f2}), then $F$ should satisfy the initial condition
  \begin{equation} \label{f7}
  F(t=0;q',q)=1.
  \end{equation}
It seems, at first sight, that it is possible to add to the
right-hand side of~(\ref{f7})
an arbitrary function of $q'-q$, which vanishes at $q'=q$. However,
this is not true. The equation for the coefficient $a_0$
  $$(q'-q) \frac{\partial a_0(q',q)}{\partial q'} =0,$$
 taken from general recursion relations for $a_n(q',q)$,
and condition $a_0(q,q)=1$ determines unambiguously
  $$a_0(q',q)=F(0;q',q)=1.$$

The problem~(\ref{f1}), (\ref{f2}), from which we have started,
has a physical
sense only for the real positive $\tau$, where $\tau=it$ (if the
heat equation and heat kernel are considered), or for real $t$ (if
the quantum mechanical evolution equation is considered). The same
restrictions
are initially fair for equation~(\ref{f6}) too. But we can
analytically continue the function $F$ into complex plain of the
variable $t$ using
the differential equation~(\ref{f6}) with condition~(\ref{f7}).
Now the variable $t$ may vary in hole complex plain $t$.
There is no restriction $\Re t >0$, which takes place for the
analytic semigroup.

If $q$ is regular point of the function $V(q)$ and at any domain
expansion in powers of $\Delta q =q'-q$ is fair
  \begin{equation} \label{f8}
  V(q') = \sum_{k=0}^{\infty} \Delta q^k \frac{V^{(k)}(q)}{k!}.
  \end{equation}
(the notation
  $$V^{(k)}(q) \equiv \frac{d^k V(q)}{dq^k}$$
is used here and will be used further everywhere),
then one can use concrete form of the coordinate dependence
of the coefficients $a_n$
  \begin{equation} \label{f9}
  F(t;q',q) = 1+
  \sum_{n=1}^{\infty} \sum_{k=0}^{\infty} (it)^n \Delta q^k b_{nk}(q).
  \end{equation}

It is obvious that
  $$\sum_{k=0}^{\infty} \Delta q^k b_{nk}(q)= a_n(q',q)=-Y_n(q',q),$$
where $Y_n$ are the functions introduced in~[Slobodenyuk 1993].
The behavior of $Y_n$ was studied in~[Slobodenyuk 1995] using
representation adduced in~[Slobodenyuk 1993].

Substitution of~(\ref{f9}), (\ref{f8}) into~(\ref{f6}) leads to
recurrent relations for the coefficients $b_{nk}$
  \begin{equation} \label{f10}
  b_{nk}=\frac{1}{n+k} \left[\frac{(k+1)(k+2)}{2} b_{n-1, k+2} -
  \sum_{m=0}^k \frac{V^{(m)}(q)}{m!} b_{n-1, k-m} \right]
  \end{equation}
with condition $b_{0k}= \delta_{k0}$. Specifically,
  \begin{equation} \label{f11}
  b_{1k}=- \frac{V^{(k)}(q)}{(k+1)!}.
  \end{equation}

Expressions~(\ref{f9}), (\ref{f10}) determine a formal solution of
problem~(\ref{f6}),~(\ref{f7}). As to the expansion in powers of
$\Delta q$ in~(\ref{f9}), one may expect that its convergence range
is equal to one for expansion~(\ref{f8}) of the potential. The
series in $t$
in~(\ref{f9}) is usually treated as divergent one. At first sight,
it is really always so. Let us estimate the convergence of the
 series in~(\ref{f9}).

At the beginning let $n$ be fixed and $k \to \infty$. Expressing
$b_{n-1, k+2}$
from~(\ref{f10}) via the coefficients with smaller $k$ we shall come
to some linear combination of the coefficients of type $b_{n_0, 0}$
and $b_{n_1, 1}$ with any indexes $n_0, n_1$ (for the sake of brevity
we shall
write further only the terms with $b_{n_0, 0}$, implying that the same
statements are concerned with the terms with $b_{n_1, 1}$). The main
growth for
large $k$ takes place if the second index of $b_{nk}$ is diminished:
a)~with using the term $V^{(k)} b_{n-1, 0}/k!$, b)~with using
the expression on the left-hand side of~(\ref{f10}).

In the case a) we get for $k \to \infty$
  \begin{equation} \label{f12}
  |b_{n-1, k+2}^{(a)}| \sim \frac{2}{(k+1)(k+2)} \frac{|V^{(k)}|}{k!}
  |b_{n-1, 0}|.
  \end{equation}
Because series~(\ref{f8}) converges at some circle with
radius $R(q)$ the estimate
  $$\frac{|V^{(k)}|}{k!} \sim \frac{1}{R^k(q)}$$
for $k \to \infty$ is fair. So, for every fixed $n$ and for
$k \to \infty$ we have
  \begin{equation} \label{f13}
  |b_{nk}^{(a)}| \sim \frac{|b_{n0}|}{R^k(q)}.
  \end{equation}
The contributions of type~(\ref{f13}) correspond to the expansion
in $\Delta q$,
which is convergent for every fixed $n$ with convergence
range $R(q)$.

In the case b) for $k \to \infty$ we get
  $$ |b_{n-1, k+2}^{(b)}| \sim \frac{2^{k/2+1} (n+k)!}{k! (n+k/2-1)!}
     |b_{n+k/2, 0}|.$$
Behavior of $b_{nk}^{(b)}$ for $k \to \infty$ depends on the behavior
 of $b_{n0}$ for $n \to \infty$. If $b_{n0}$ decreases when
 $n \to \infty$ or
increases more slowly than $\Gamma(\alpha n)$ ($\alpha$ is any
positive number), then
  $$|b_{nk}^{(b)}| \sim \frac{|b_{n0}|}{\Gamma (k/2)}$$
for $k \to \infty$, i.e., these contributions will disappear at
large $k$.
If $b_{n0}$ increases as $\Gamma(\alpha n)$ (here $0< \alpha \le 1$,
in~[Slobodenyuk 1995] showed that $\alpha$ cannot be larger then 1),
then for $k \to \infty$ and $\alpha < 1$
  $$|b_{nk}^{(b)}| \sim
       \frac{|b_{n0}|}{\Gamma \left( \frac{1-\alpha}{2}k \right)},$$
so, these contributions will disappear too with the
growth of $k$. If $\alpha =1$,
then the following estimate will take place ($n$ is fixed,
$k \to \infty$)
  \begin{equation} \label{f15}
  |b_{nk}^{(b)}| \sim |b_{n0}| k^c \rho ^k.
  \end{equation}
In this case the expansion in $\Delta q$ in~\ref{f9}) will have the
finite convergence range too, but it will be equal to minimum from
two values $R(q)$ and $\rho$.

Now let us examine the behavior of $|b_{n0}|$ (the same will be also
correct for
$|b_{n1}|$) when $n \to \infty$. Consider the decreasing of $n$ till
1 by means of the first term on the right-hand side of~(\ref{f10})
  \begin{equation} \label{f16}
  |b_{n0}| \sim \frac{|b_{n-1, 2}|}{n} \sim \cdots \sim
  \frac{(n-1)!}{2^{n-1}} |b_{1,2n-2}|=
  \frac{(n-1)!}{2^{n-1}} \frac{|V^{(2n-2)}|}{(2n-1)!}.
  \end{equation}
Because $|V^{(k)}| \sim k!/R^k(q)$ for $k \to \infty$, then
for $n \to \infty$ we get
  \begin{equation} \label{f17}
  |b_{n0}| \sim \frac{(n-1)!}{2^{n-1} (2n-1)} \sim n!.
  \end{equation}

Really, the contributions taken into account in~(\ref{f16})
provide the main growth only for the potentials, for which
$R(q)<\infty$. If the potentials with $R(q)=\infty$ are
considered (e.g., polynomial ones), then, at first sight, one can
conclude
from~(\ref{f16}) that $|b_{n0}| \sim 1/n!$. But it is not so, in fact.
As it was shown in~[Slobodenyuk 1995], the combination of
contributions of the first
term and terms of sum over $m$ in~(\ref{f10}) leads to the estimate of
 type $|b_{n0}| \sim \Gamma(\alpha n)$.

So, for arbitrary potentials the series in $t$ in~(\ref{f9}) is
divergent.
But in our estimates, in fact, absolute values of all contributions to
every coefficient $b_{nk}$ were summed. Nevertheless, for some
potentials
the cancellation of different terms may occur. It can lead to
convergence of
the expansion in~(\ref{f9}). For the potentials considered in
Secs.~3--5 this
cancellation takes place only for definite values of the coupling
constant.

Note that we, really, test expansion~(\ref{f9}) for the absolute
convergence.
So, it is enough for the convergence of double series that~(\ref{f9}),
in which
instead of $b_{nk}$ absolute values $|b_{nk}|$ taken, would converge
for any
order of summation. Our consideration corresponds to the following
order: at
first the series over $k$ for every fixed n are summed and then
summation
over $n$ is made. If one assumes that there is convergence of the
series
in index $n$ then, as it was shown before, the convergence in index
$k$ will
take place at every fixed $n$, and to establish the convergence of
the series
in index $n$ it is enough to determine the behavior  of the
coefficients $b_{n0}, \ b_{n1}$ only (but not all $b_{nk}$) at
$n \to \infty$.

\section{Modified P\"oschl --- Teller potential}

Let us introduce standard notation for the coupling constant $g=
\lambda (\lambda -1)/2$ ($\lambda >0$) and investigate
modified P\"oschl --- Teller potential
  \begin{equation} \label{f3.1}
  V(q)= - \frac{\lambda (\lambda -1)}{2} \frac{\beta^2}
  {\cosh ^2(\beta q)}
  \end{equation}
for the convergence of expansion~(\ref{f9}).

Because the constant $\beta$ is connected with the choice of length
scale one can put $\beta=1$ without the restriction of generality.
Further, for the sake of brevity we shall denote
  \begin{equation} \label{f3.2}
  f(q)= - \frac{1}{\cosh ^2q}.
  \end{equation}
Then the potential reads briefly $V(q)=gf(q)$.

The potential~(\ref{f3.1}) has the expansion of type~(\ref{f8}) about
every real  point $q$. Its convergence range is equal to $R(q)=
\sqrt{(\pi/2)^2+q^2}$
and is determined by the distance to the nearest singularities of the
function $1/\cosh ^2 q$ placed at the points $q= \pm i \pi/2$.
 The derivatives can be calculated as follows
  \begin{equation} \label{f3.3}
  V^{(k)}(q)= gf^{(k)}(q),
  \end{equation}
where $f^{(k)}$ are represented as expansions in powers of $f$
  \begin{equation} \label{f3.4}
  f^{(2n)}(q)= \sum_{l=1}^{n+1} a_l^{(2n)} f^l(q),
  \end{equation}
  \begin{equation} \label{f3.5}
  f^{(2n+1)}(q)= \sum_{l=1}^{n+1} la_l^{(2n)} f^{l-1}f^{(1)}=
                 \sum_{l=1}^{n+1} a_l^{(2n+1)} f^{l-1}f^{(1)}.
  \end{equation}

To obtain all coefficients $a_l^{(k)}$ it is enough to put $a_l^{(0)}=
\delta_{l1}$ and take into account
  $$(f^{(1)})^2=4f^3+4f^2.$$
For $a_l^{(2n)}$ one has the recursion relations
  \begin{equation} \label{f3.6}
  a_l^{(2n)}= 4l^2 a_l^{(2n-2)} +4(l-1)(l-1/2) a_{l-1}^{(2n-2)}.
  \end{equation}
So, every derivative of the function $f(q)$ is represented as a
polynomial in powers of this function.

From~(\ref{f10}) one gets for potential~(\ref{f3.1})
  \begin{equation} \label{f3.7}
  b_{nk}=\frac{1}{n+k} \left[\frac{(k+1)(k+2)}{2} b_{n-1, k+2} -
  \frac{\lambda (\lambda -1)}{2} \sum_{m=0}^k
  \frac{f^{(m)}}{m!} b_{n-1, k-m} \right],
  \end{equation}
where the derivatives $f^{(m)}$ are calculated
via~(\ref{f3.4})--(\ref{f3.6}).

According to the note at the end of Sec.~2, it is enough
for testing the convergence of series~(\ref{f9})
to examine the behavior at $n \to \infty$ of the coefficients
$b_{n0}, \ b_{n1}$ only. Introduce in this connection the functions
  \begin{equation} \label{f3.8}
  B_k(t,q)= \sum_{n=0}^{\infty} t^n b_{nk}(q)
  \end{equation}
and consider them for $k=0,\ 1$.

The analysis of relations~(\ref{f3.7}) with taking into account
of~(\ref{f3.4})--(\ref{f3.6}) shows that $B_0, \ B_1$ can be
represented in the form
  \begin{eqnarray} \label{f3.9}
  B_0(t,q)&=&1+ \sum_{n=1}^{\infty} (it)^n \sum_{l=1}^n
           \frac{(-1)^l}{l!} f^l(q) \beta_{nl}
           \prod_{j=1}^l \left(\frac{\lambda (\lambda -1)}{2}
           - \frac{j(j-1)}{2} \right)
  \nonumber \\ &=&
           1+ \sum_{n=1}^{\infty} (it)^n \sum_{l=1}^n
           \frac{(-1)^l}{l!} f^l(q) \beta_{nl}
           \frac{\Gamma(\lambda+l)}{2^l \Gamma(\lambda-l)},
  \end{eqnarray}
  \begin{eqnarray} \label{f3.10}
  B_1(t,q)&=& \sum_{n=1}^{\infty} (it)^n \sum_{l=1}^n
           \frac{(-1)^l}{l!} \frac{l}{2} f^{l-1}(q) f^{(1)}(q)
           \beta_{nl}
           \prod_{j=1}^l \left(\frac{\lambda (\lambda -1)}{2}
           - \frac{j(j-1)}{2} \right)
  \nonumber \\ &=&
            \sum_{n=1}^{\infty} (it)^n \sum_{l=1}^n
           \frac{(-1)^l}{l!} \frac{l}{2} f^{l-1}(q) f^{(1)}(q)
           \beta_{nl}
           \frac{\Gamma(\lambda+l)}{2^l \Gamma(\lambda-l)},
  \end{eqnarray}
where
  \begin{equation} \label{f3.11}
  \beta_{nl}= \frac{1}{2^{n-l}} \frac{(n-1)!}{(l-1)!}
                 \frac{a_l^{(2n-2)}}{(2n-1)!}.
  \end{equation}

To estimate the behavior of $\beta_{nl}$ when $n \to \infty$ we
probe the asymptotics of $a_l^{(2n-2)}$. Let us take $a_l^{(2n-2)}$
for sufficiently
large $n$ and begin to express $a_l^{(2n-2)}$ with the help
of~(\ref{f3.6}) via
the coefficients with smaller $n$ and $l$ so that to come to
$a_1^{(0)}=1$
at the end. Maximal contribution arises in this procedure when,
at the beginning, $n$ will be diminished at fixed $l$ by means of
the first term
on the right-hand side of~(\ref{f3.6}), and then, when the
equality $n=l-1$
becomes valid, $n$ and $l$ will start to be decreased simultaneously
by unit at
every step by means of the second term in~(\ref{f3.6}). For large $n$
this gives the estimate
  $$a_l^{(2n-2)} \sim 4^{n-l} l^{2(n-l)} (2l-1)!.$$
Then $\beta_{nl}$ behaves itself as
  \begin{equation} \label{f3.12}
  \beta_{nl} \sim 2^{n-l} l^{2(n-l)} \frac{(n-1)!}{(l-1)!}
                 \frac{(2l-1)!}{(2n-1)!}.
  \end{equation}

Now one can evaluate the asymptotics at $n \to \infty$ of the
coefficients of series~(\ref{f3.9}), (\ref{f3.10}).
Taking in~(\ref{f3.9}), (\ref{f3.10}) in the sum over $l$ the term
with $l=n$ we shall obtain for noninteger $\lambda$
that the coefficients in front of $t^n$ growth
in~(\ref{f3.9}) as
  $$\frac{f^n}{2^n} \frac{\Gamma(\lambda+n)}
                            {n! \Gamma(\lambda-n)} \sim n!,$$
and in~(\ref{f3.10}) as
  $$\frac{f^{n-1}f^{(1)}}{2^{n+1}} \frac{\Gamma(\lambda+n)}
                        {(n-1)! \Gamma(\lambda-n)} \sim n!,$$
So, for noninteger $\lambda$ series~(\ref{f3.9}),~(\ref{f3.10}),
and, hence, (\ref{f9}) are asymptotic ones.

Let now $\lambda$ be integer ($\lambda >1$). Then
in~(\ref{f3.9}),~(\ref{f3.10}) in the sums over $l$ only the terms with
$l< \lambda$ are different from zero, and, in fact, one should take
instead of
$\sum_{l=1}^n$ the sum $\sum_{l=1}^{\min \{n,\lambda-1\} }$.
For $n \ge
\lambda -1$ the sum over $l$ will always contain $\lambda -1$
terms, and its dependence on $n$ will be determined only
by the dependence on $n$ of the coefficients $\beta_{nl}$.
And the dependence of the latter on $n$,
as it is clear from estimate~(\ref{f3.12}), at fixed
$l \le \lambda -1$ and
at $n \to \infty$ is determined by the factor
  $$\beta_{nl} \sim \left(2(\lambda -1)^2 \right)^n
  \frac{(n-1)!}{(2n-1)!}.$$
So, the coefficients in front of $t^n$ in~(\ref{f3.9}),~(\ref{f3.10})
behave themselves at large $n$ as
  $$\frac{C^n (n-1)!}{(2n-1)!}$$
with any positive $C$,
i.e., the series will be convergent at the circle of infinite range.

To obtain finally the function $F(t;q',q)$ it is necessary either to
 take the
coefficients $b_{n0}, \ b_{n1}$ from~(\ref{f3.9}),~(\ref{f3.10}) to
calculate
other $b_{nk}$ using~(\ref{f3.7}), or starting from $B_0, \ B_1$ to
calculate
other functions $B_k(t,q)$ from the equations
  \begin{equation} \label{f3.13}
  B_{k+2}=\frac{2}{(k+1)(k+2)} \left( \frac{1}{i}
          \frac{\partial B_k}{\partial t} +
    \frac{k}{it} B_k + g \sum_{m=0}^k \frac{f^{(m)}}{m!}
    B_{k-m} \right),
  \end{equation}
which in an obvious way are derived from~(\ref{f6}) after
substitution
  \begin{equation} \label{f3.14}
  F(t;q',q)= \sum_{k=0}^{\infty} \Delta q^k B_k(t,q),
  \end{equation}
and substitute them into~(\ref{f3.14}).

Particularly, for $\lambda =2 \ (g=1)$ we have the potential $V(q)=
-1/\cosh^2 q$, for which
  \begin{equation} \label{f3.15}
  B_0(t,q)=1- f(q) \sum_{n=1}^{\infty} \frac{(it)^n}{(2n-1)!!}=
           1- f(q) \sqrt{\frac{\pi it}{2}} e^{it/2} \erf (\sqrt{it/2}),
  \end{equation}
  \begin{equation} \label{f3.16}
  B_1(t,q)=-\frac{1}{2} f^{(1)}(q) \sum_{n=1}^{\infty}
  \frac{(it)^n}{(2n-1)!!}=
   -\frac{1}{2} f^{(1)}(q) \sqrt{\frac{\pi it}{2}} e^{it/2}
   \erf (\sqrt{it/2}).
  \end{equation}
With the help of~(\ref{f3.13}) one is able to determine all
coefficient functions
$B_k$ starting from~(\ref{f3.15}),~(\ref{f3.16}) and then to substitute
them into~(\ref{f3.14}). In this manner the function $F$ will be found.

We established, that for integer $\lambda$ expansion~(\ref{f9}) was
convergent if $|\Delta q| < R(q)$ and the representation~(\ref{f4}),
~(\ref{f9}) for the evolution operator kernel was not asymptotic. The
 function
$F$ is single-valued analytic in the entire complex plain of the
variable $t$ function and it has an essential singularity at the
infinite $(t=\infty)$ point.

The potential~(\ref{f3.1}) is the representative of class
of potentials studied in~[Osborn 1983]. It can be written as the
Fourier transform
  $$V(x)= \int\limits_{-\infty}^{+\infty} e^{i \alpha x}
                                             d \mu (\alpha),$$
where
  $$d \mu (\alpha) = - \frac{g \alpha d \alpha}
                                   {2 \sinh (\pi \alpha/2)}.$$
In~[Osborn 1983] showed that $|a_n| < n^{2n}/n! \sim n!$ when
$n \to \infty$ for the potentials of that class. This
does not mean that the Schwinger --- DeWitt expansion
should be divergent in every case, because this is only
bound from up, but not from down. So, our result about
convergence of the expansion for the potential~(\ref{f3.1})
for integer $\lambda$ does not contradict to conclusions
of paper~[Osborn 1983].

\section{Potential $V(q)=g/q^2$}

Another example of convergent series~(\ref{f9}) we shall get
considering the potential
  \begin{equation} \label{f4.1}
  V(q)= \frac{\lambda (\lambda -1)}{2} \frac{1}{q^2}
  \end{equation}
on half-line $q>0$. This potential is well studied and it is
known analytic expression for the kernel for it. Our purpose
is to show  how the method described above can be applied
to singular potentials.

Expansion~(\ref{f8}) for the potential~(\ref{f4.1}) has
the finite convergence range $R(q)=q$, finiteness of which
is connected with singularity of $V(q)$ at the point $q=0$.
The derivatives $V^{(k)}$ may be easily calculated
  \begin{equation} \label{f4.2}
  V^{(k)}(q)= (-1)^k \frac{\lambda (\lambda -1)}{2}
  \frac{(k+1)!}{q^{k+2}}.
  \end{equation}
But for this potential additional problem arises because of
its singularity at the origin. To obtain the kernel that provides
fulfilment of boundary condition for the wave function $\psi (q)$
at $q=0$ ($\psi(q)$ should vanish at $q=0$) one is to use initial
condition of more general form as compared with~(\ref{f2}).
Namely, in this case
  \begin{equation} \label{f4.3}
  \langle q',t=0\mid q,0 \rangle = \delta (q'-q)+A\delta(q'+q),
  \end{equation}
where constant $A$ is determined by demand that the kernel does
not have singularity at $q=0$ and/or $q'=0$ ($t \ne 0$).
In correspondence with~(\ref{f4.3}) and analogously to~(\ref{f4})
the kernel may be represented as
  \begin{eqnarray} \label{f4.4}
  \langle q',t\mid q,0 \rangle &=& \frac{1}{\sqrt{2\pi it}}
  \exp \left\{i \frac{(q'-q)^2}{2t} \right\} F^{(-)}(t;q',q)
  \nonumber \\ &+&
  A \frac{1}{\sqrt{2\pi it}}
  \exp \left\{i \frac{(q'+q)^2}{2t} \right\} F^{(+)}(t;q',q).
  \end{eqnarray}
The equations for the functions $F^{(\pm)}$ are
  \begin{equation} \label{f4.5}
  i \frac{\partial F^{(\pm)}}{\partial t} =
  -\frac{1}{2} \frac{\partial ^2 F^{(\pm)}}{\partial q'^2} +
  \frac{q'\pm q}{it} \frac{\partial F^{(\pm)}}{\partial q'}
  + V(q')F^{(\pm)},
  \end{equation}
and initial conditions are
  \begin{equation} \label{f4.6}
  F^{(\pm)}(t=0;q',q)=1.
  \end{equation}
Directly in~(\ref{f4.5}), (\ref{f4.6}) $q',\; q >0$. But it is
possible to consider analytic continuation of $F^{(\pm)}$
into the complex plain $q$ and adopt negative values of $q$. Then
one may write
  \begin{equation} \label{f4.7}
  F^{(+)}(t;q',q)=F^{(-)}(t;q',-q)
  \end{equation}
and study only one function $F(t;q',q)=F^{(-)}(t;q',q)$,
where $q', \; q$ may be both positive and negative.

It is convenient to begin with positive $q', \; q$. In this
case technique described above can be used without any changes.
From~(\ref{f10}) with account of~(\ref{f4.2}) we take
  \begin{equation} \label{f4.8}
  b_{nk}=\frac{1}{n+k} \left[\frac{(k+1)(k+2)}{2} b_{n-1, k+2} +
  \frac{\lambda (\lambda -1)}{2} \sum_{m=0}^k (-1)^{m+1}
  \frac{m+1}{q^{m+2}} b_{n-1, k-m} \right].
  \end{equation}
If we shall diminish $n$ times the first index of $b_{nk}$ by means
of~(\ref{f4.8}), then we get
  \begin{eqnarray} \label{f4.9}
  b_{nk}&=& \frac{(-1)^{n+k}}{q^{2n+k}} \frac{(k+n-1)!}{n!(n-1)!k!}
        \prod_{j=1}^n \left(\frac{\lambda (\lambda -1)}{2}
        - \frac{j(j-1)}{2} \right)
  \nonumber \\ &=&
        \frac{(-1)^{n+k}}{q^{2n+k}} \frac{(k+n-1)!}{n!(n-1)!k!}
        \frac{\Gamma(\lambda+n)}{2^n \Gamma(\lambda -n)}.
  \end{eqnarray}
It is obvious that for noninteger $\lambda$
$|b_{nk}| \sim n!$ when $n \to \infty$.
So, for noninteger $\lambda$ expansion~(\ref{f9})
for potential~(\ref{f4.1}) is divergent. But if $\lambda$ is
integer ($\lambda >1$, because cases
$\lambda =0, \ \lambda =1$ are trivial) then one can easy
see from~(\ref{f4.9}) that only the coefficients $b_{nk}$ for
$n< \lambda$ are
different from zero, and in~(\ref{f9}) the series in powers of $t$ is
really the polynomial of finite degree $\lambda -1$.

Let us substitute~(\ref{f4.9}) into~(\ref{f9}) and
make summation over $k$. Then we get finally
  \begin{equation} \label{f4.11}
  F(t;q',q)= 1+ \sum_{n=1}^{\infty}
             \left( \frac{-it}{2q'q} \right)^n
             \frac{\Gamma(\lambda+n)}{n! \Gamma(\lambda-n)}.
  \end{equation}
Derivation of~(\ref{f4.11}) is made in supposition
that $|\Delta q| < |q|$. If this condition is not satisfied, then
calculations should be made with the expansion about the point
$q'$ in powers of $q-q'$. But because
$F$ is symmetric in $q', \ q$, then it is clear that the answer
in this case would be the same as in~(\ref{f4.11}).
Note, that representation~(\ref{f4.11}) for $F$ does not suppose
that $q', \ q>0$. One can put $q'$ and/or $q$ negative and,
hence, (\ref{f4.11}) gives expressions both for $F^{(-)}$
and for $F^{(+)}$.

Expansion~(\ref{f4.11}) has singularity at $q=0$ ($q'=0$). For
noninteger $\lambda$ this expression is asymptotic, so it
cannot be applied to analysis the behaviour of $F$ at
$q',\, q \to 0$.
It is correct only for sufficiently small values of variable
$t/q'q$. But for integer $\lambda$~(\ref{f4.11}) becomes
finite series because in this case $1/\Gamma(\lambda -n)=0$
for $n \ge \lambda$ and sum over $n$ should be made only
till $\lambda -1$, but not till infinity. One can state, so,
that $F^{(\pm)}$ is really singular at the point $q=0$ (or
$q'=0$). This feature, nevertheless, is not so dangerous,
because only the kernel~(\ref{f4.4}) should be finite at
$q=0$ ($q'=0$).

After substitution of~(\ref{f4.11}) into~(\ref{f4.4}) we get
  \begin{eqnarray} \label{f4.12}
  \langle q',t\mid q,0 \rangle &=& \frac{1}{\sqrt{2\pi it}}
  \exp \left\{ i \frac{q'^2+q^2}{2t} \right\}   \left\{
  e^{-i \frac{q'q}{t}}
  \sum_{n=0}^{\lambda -1} \left( \frac{-it}{2q'q} \right)^n
          \frac{\Gamma(\lambda+n)}{n! \Gamma(\lambda-n)}  \right.
  \nonumber \\ &+&
  \left. A e^{i \frac{q'q}{t}}
  \sum_{n=0}^{\lambda -1} \left( \frac{it}{2q'q} \right)^n
          \frac{\Gamma(\lambda+n)}{n! \Gamma(\lambda-n)} \right\}.
  \end{eqnarray}
Expanding $\exp\{ \pm i q'q/t \}$ into series in $q'q/t$ and
considering terms with singularity in variable $q'q$ we
can see that if
$A=e^{-i \pi \lambda}$ then all these terms will be cancelled
and the kernel will be equal to zero when $q'q=0$ (this will
be zero of order $\lambda$). So, the initial condition~(\ref{f4.3})
with $A=e^{-i \pi \lambda}$ provides fulfillment of the boundary
condition for the kernel at origin.

Finally the kernel may be represented as
  \begin{eqnarray} \label{f4.13}
  \langle q',t\mid q,0 \rangle &=& \frac{1}{\sqrt{2\pi it}}
  \exp \left\{ i \frac{(q'-q)^2}{2t} \right\}
  \sum_{n=0}^{\infty} \left( \frac{-it}{2q'q} \right)^n
          \frac{\Gamma(\lambda+n)}{n! \Gamma(\lambda-n)}
  \nonumber \\ &+&
  e^{-i \pi \lambda} \frac{1}{\sqrt{2\pi it}}
  \exp \left\{ i \frac{(q'+q)^2}{2t} \right\}
  \sum_{n=0}^{\infty} \left( \frac{it}{2q'q} \right)^n
          \frac{\Gamma(\lambda+n)}{n! \Gamma(\lambda-n)}.
  \end{eqnarray}
For integer $\lambda$ sums are made only till $\lambda -1$
and this expansion is finite, for noninteger $\lambda$ it
is asymptotic.

Naturally, this result exactly coincides with well known
representation
  \begin{eqnarray} \label{f4.14}
  &&\langle q',t\mid q,0 \rangle = e^{-i\frac{\pi}{2}(\lambda-1/2)}
  \frac{\sqrt{q'q}}{it}
  \exp \left\{i \frac{q'^2+q^2}{2t}  \right\}
  J_{\lambda-1/2}\left( \frac{q'q}{t}\right)
  \nonumber \\
  &&= \frac{1}{\sqrt{2\pi it}}
  \exp \left\{i \frac{(q'-q)^2}{2t} \right\}
  \sqrt {\frac{\pi q'q}{2it}} e^{-i\frac{\pi}{2}(\lambda-1/2)}
  e^{\frac{iq'q}{t}}
  H_{\lambda-1/2}^{(2)} \left( \frac{q'q}{t}\right)
    \nonumber \\
  &&+ e^{-i\pi \lambda} \frac{1}{\sqrt {2\pi it}}
  \exp \left\{i \frac{(q'+q)^2}{2t} \right\}
  \sqrt {\frac{\pi q'q}{2it}} e^{i\frac{\pi}{2}(\lambda+1/2)}
  e^{-\frac{iq'q}{t}}
  H_{\lambda-1/2}^{(1)} \left( \frac{q'q}{t}\right),
  \end{eqnarray}
which may be derived directly by reducing the Schr\"odinger
equation to the equation for the cylindric functions
(here $J_\nu$ is Bessel function, $H_\nu^{(1,2)}$ are Hankel
functions of first and second kinds). Strictly speaking,
validness of~(\ref{f4.13}) for noninteger $\lambda$ follows
only from~(\ref{f4.14}), but not from previous considerations.

Note, that essential feature of representation~(\ref{f4.13})
for the kernel is following: the sums over index $n$ are
divergent if $\lambda$ is noninteger and, contrary, they are
finite if $\lambda$ is integer. The kernel is well defined
in both cases and apparent expression~(\ref{f4.14}) allows
us to study its behaviour in different variables: $t, \ q', \
q$ or $\lambda$. But if we do not know exact solution, as it
usually takes place in more complicated problems with
other potentials, and if we have only asymptotic expansion
of the form~(\ref{f4.4}), then we will have many problems for
noninteger $\lambda$ and much less problems for integer
$\lambda$.

For example, for the potential~(\ref{f4.1}) we cannot study
from the asymptotic expansion~(\ref{f4.13}) behaviour of the
kernel at $q \to 0$. Particularly, if we had not exact
expression~(\ref{f4.14}), but had only asymptotic
one~(\ref{f4.13}), we would not know that the kernel has the
zero of order $\lambda$ at $q'q \to 0$.

From the other side, if we try to get the expansion in powers
of the coupling constant $g=\lambda (\lambda -1 )/2$ for $F$
starting from~(\ref{f4.11}), we would get after some
transformations
  \begin{equation} \label{f4.15}
  F(t;q',q)= 1+ \sum_{k=1}^{\infty} g^k \sum_{n=k}^{\infty}
          \left( \frac{-it}{q'q} \right)^n \frac{C_{nk}}{n!},
  \end{equation}
where $C_{nk}$ are the coefficients of the polynomial
  \begin{equation} \label{f4.16}
  \prod_{j=1}^n \left( g- \frac{j(j-1)}{2} \right) =
  \sum_{k=1}^n g^k C_{nk}.
\end{equation}
Equation~(\ref{f4.15}) is the series of the conventional
perturbation theory. This series is not simply divergent,
but its coefficients are divergent too. Consider contribution
of the first order in $g$:
  $$ F^{(1)}(t;q',q)= g \sum_{n=1}^{\infty}
         \left( \frac{-it}{q'q} \right)^n \frac{C_{n1}}{n!}. $$
From~(\ref{f4.16}) one can easy derive
  $$ C_{n1}= (-1)^{n-1} \frac{n! (n-1)!}{2^{n-1}}. $$
Hence,
  \begin{equation} \label{f4.17}
  F^{(1)}(t;q',q)= -2g \sum_{n=1}^{\infty}
          \left( \frac{-it}{2q'q} \right)^n (n-1)!,
  \end{equation}
and the coefficient in front of $g$ is divergent series.
Now we see, that without knowledge of exact solution for
noninteger $\lambda$ we cannot build even conventional
perturbation theory for $F$. Nevertheless, for integer $\lambda$
the Schwinger --- DeWitt expansion is convergent and it
can be used for further applications.

\section{Other samples of potentials}

Calculations made in Secs.~3,~4 may be easily repeated
for some similar potentials which are often used in
one-dimensional many-body problems~[Olshanetsky 1983, Calogero 1975,
Sutherland 1971, 1972].
These are the potentials
  \begin{equation} \label{f5.1}
  V(q)= \frac{\lambda (\lambda -1)}{2} \frac{1}{\sinh ^2q},
  \end{equation}
and
  \begin{equation} \label{f5.2}
  V(q)= \frac{\lambda (\lambda -1)}{2} \frac{1}{\sin ^2q}.
  \end{equation}
To prove convergence of the series for $F^{(-)}(t;q',q) \,
(q',\ q>0)$ it is enough to make a little modification of
considerations of Sec.~3.

For the potential~(\ref{f5.1}) denote
  \begin{equation} \label{f5.3}
  f(q)= \frac{1}{\sinh ^2q},
  \end{equation}
and notice, that
  $$(f^{(1)})^2=4f^3+4f^2,$$
i.e., it exactly coincides with the
corresponding expression for the function $f(q)$ defined
by~(\ref{f3.2}) in Sec.~3. So, all relations for the
derivatives of $f$ obtained there and,
hence, expressions for $b_{nk}, \ B_k$, and $F$ (now it is
$F^{(-)}$) remain right in this case.
There exist only two differences: function $f$ is defined now
by~(\ref{f5.3}), but not by~(\ref{f3.2}), and convergence
range of the expansion~(\ref{f8}) is $R(q)= \sqrt{\pi^2+q^2}$.

Hence, convergence of~(\ref{f5}) for the potential~(\ref{f5.1})
when $q', \, q >0$ takes place for integer $\lambda$. The
function $F^{(-)}$ is single-valued and analytic function in the
entire complex plain of the variable $t$ for all $q', \, q >0$.

The potential~(\ref{f5.2}) in region $0<q<\pi$
can be also considered in similar way. Denote
  \begin{equation} \label{f5.4}
  f(q)= \frac{1}{\sin ^2q}
  \end{equation}
and take into account, that
  $$(f^{(1)})^2=4f^3-4f^2.$$
This expression differs from analogous ones for
potentials~(\ref{f3.1}),~(\ref{f5.1}) only by the sign of
the second term. So, we are able to reconstruct
the expressions from Sec.~3 with small changes only:
in~(\ref{f3.9}),~(\ref{f3.10}) will appear an additional
multiplier $(-1)^{n+l}$, and the function $f$ will be
defined by~(\ref{f5.4}).

Conclusion about convergence of the expansion~(\ref{f5})
for $F^{(-)}$ at region $0<q', \, q<\pi$ when $\lambda$
is integer remains fair for the potential~(\ref{f5.2}).
But both these potentials are singular at $q=0$. So, one
is to consider initial condition~(\ref{f4.3}) and additional
function $F^{(+)}(t;q',q)$ as it was made in Sec.~4. Really
it is enough to continue $F^{(-)}$ into region $q<0$ and
use~(\ref{f4.6}). Expressions obtained in Sec.~3 do not
allow us to make any conclusions about behaviour of $F$
for $q<0$.

Let us consider another representation for $a_n(q',q)$. The
potentials~(\ref{f5.1}),~(\ref{f5.2}) may be written as
follows
  \begin{equation} \label{f5.5}
  V(q)= g \left( \frac{1}{q^2} +
                       \sum_{k=0}^{\infty} s_k q^k \right).
  \end{equation}
where $s_k$ are known coefficients, $g=\lambda (\lambda -1)/2$.
The coefficient functions $a_n(q',q)$ have a form of Loran
series with finite number of pole terms
  \begin{equation} \label{f5.6}
  a_n(q',q)= \sum_{k=-n}^{\infty} \sum_{l=-n}^{\infty}
                                            q'^k q^l d^n_{kl}.
  \end{equation}
Substitution of~(\ref{f5}), (\ref{f5.5}), (\ref{f5.6})
into~(\ref{f4.4}) gives us algebraic recurrent relations for
$d^n_{kl}$
  \begin{equation} \label{f5.7}
  (n+k)d^n_{kl} \pm (k+1)d^n_{k+1,l} =
       \left( \frac{(k+1)(k+2)}{2} -g \right) d^{n-1}_{k+2,l}
       -g \sum_{m=k+n-1}^{\infty} s_m d^{n-1}_{k-m,l}.
  \end{equation}
If $k+n-1 <0$ then sum at last term of~(\ref{f5.7}) should
be equated to zero. One can find solutions of
equation~(\ref{f5.7}) and be convinced of validness of
representation~(\ref{f5.6}). It is obvious, besides, that the
most singular at $q', \, q =0$ terms in $a_n$ do not depend
on $s_m$ and they exactly coincide with ones for the
potential~(\ref{f4.1}).

We know that $F^{(-)}$ ($q', \, q>0$) for considered
potentials is represented by convergent series of type~(\ref{f9}).
Then there exist representation of type~(\ref{f5}), (\ref{f5.6})
and Loran series~(\ref{f5.6}) is convergent at pierced
in zero polycircle of finite radii. At this series sign
of $q$ has no any meaning. One may consider both $q>0$ and
$q<0$. Convergence will take place in any case.

One can state that convergence of expansion~(\ref{f5})
for $F^{(+)}$ follows from convergence of it for $F^{(-)}$.
The latter was proved earlier. So, we established that for the
potentials~(\ref{f5.1}),~(\ref{f5.2}) the Schwinger ---
DeWitt expansion~(\ref{f4.4}) is convergent for integer
$\lambda$. Singularities of $F^{(\pm)}$ at $q', \, q=0$
cancel each other in combination~(\ref{f4.4}) if
$A=e^{-i\pi \lambda}$ so as in the case of the
potential~(\ref{f4.1}).

\section{Conclusion}

Usually the Schwinger --- DeWitt expansion is used as
asymptotic. Its general property is rising of the
coefficients $a_n(q',q)$ as $n!$ for $n \to \infty$
(or as $\Gamma \left( \frac{L-2}{L+2}n \right)$ if the
potential is polynomial of order $L$)~[Osborn 1983, Slobodenyuk 1995].
Such growth takes place always when no any cancellations
of different contributions into $a_n$ occur. It is so for the
most number of potentials. But there exist some potentials,
for which cancellations really occur.
Examples of such potentials were considered in present
paper. It was proved convergence of the Schwinger ---
DeWitt expansion for them when constant $\lambda$ is
integer and divergence when $\lambda$ is noninteger.

Besides the potentials mentioned above one more example is
known~[Slobodenyuk 1996~\cite{TMF2}], which has the property of convergence
of the expansion~(\ref{f9}). It is following:
  \begin{equation} \label{f6.1}
  V(q)=a^2q^2 + \frac{\lambda(\lambda-1)}{2} \frac{1}{q^2}.
  \end{equation}
The expansion~(\ref{f5}) converges for it when $\lambda$
is integer. But convergence range is finite contrary to examples
of this paper. It is naturally, because expansion for the
harmonic oscillator $V(q)=a^2q^2$ has finite convergence
range. And it is so for the perturbed oscillator~(\ref{f6.1}) too.

So, we discovered existence of the class of nontrivial
potentials in the quantum mechanics, for which the
Schwinger --- DeWitt expansion is convergent and for
which initial condition for the evolution problem is
fulfilled in rigorous (analytic) sense, but not
only asymptotically (when $F$ is not analytic at $t=0$
and its value at origin is determined from condition
of continuity).

The potentials belonging to this class has at least
two remarkable features: 1) the Schwinger --- DeWitt
expansion is convergent for them, hence,
other expansions, which may be derived from it are
convergent too, and many problems, connected with
divergence of the expansions are absent for such potentials,
2) the potentials of this class have discrete coupling
constants that corresponds to discreteness of the charge
in the nature.

This is why the potentials from this class is to be well
studied. But research of quantum mechanical models is
only preparation for practical using of discovered phenomenon.
One can expect to construct the fundamental theory of
elementary particles as a result of transmission of this
approach  into the quantum field theory. It is possible to
introduce interaction in the field theory which is
analogous in any meaning to the quantum mechanical potentials
studied in this article. One of such quantum field models is
under consideration at present time and will be described
in consequent paper. The model conserves essential features
discussed above.

\newpage

\end{document}